\documentclass[journal]{IEEEtran}
\ifCLASSINFOpdf
\else
\fi
\usepackage{amsmath,graphicx}

\usepackage{setspace}
\usepackage{amsthm}
\usepackage{amssymb}
\usepackage{amsfonts}
\usepackage{color}
\usepackage{epstopdf}
\usepackage{subfigure}  

\usepackage{mathtools}
\usepackage{stmaryrd}
\usepackage[mathscr]{euscript}
\usepackage{mathrsfs}

\usepackage{soul}
\usepackage{arydshln}
\usepackage{mathtools, cuted}
\usepackage{stfloats}
\usepackage{xcolor}
\usepackage{stackengine}
\def\delequal{\mathrel{\ensurestackMath{\stackon[1pt]{=}{\scriptstyle\Delta}}}}
\usepackage{mathtools}
\usepackage{url}

\usepackage{hyperref}
\hypersetup{
	colorlinks=false,
	linkcolor=blue,
	filecolor=magenta,      
	urlcolor=blue,
}

\usepackage{algpseudocode}
\usepackage{algorithm}
\usepackage{xspace}

\DeclareMathOperator*{\argmin}{argmin}
\DeclareMathOperator{\prox}{prox}
\newcommand{\restartnow}{{\small \texttt{restart-now}}\xspace}
\newcommand{\restartallowed}{{\small \texttt{restart-allowed}}\xspace}

\begin{document}

%
\title{Autotuning Plug-and-Play Algorithms for MRI}
%
%
%

\author{Saurav K. Shastri$^1$, Rizwan Ahmad$^{1,2}$, and Philip Schniter$^1$\thanks{This work was supported in part by NIH-R01HL135489.} \\ $^1$ Dept. of Electrical and Computer Engineering, The Ohio State University, Columbus, OH, 43210.\\
	$^2$ Dept. of Biomedical Engineering, The Ohio State University, Columbus, OH, 43210.\\Email: \{shastri.19,schniter.1,ahmad.46\}@osu.edu}
\maketitle
\thispagestyle{empty}
\begin{abstract}
For magnetic resonance imaging (MRI), recently proposed ``plug-and-play'' (PnP) image recovery algorithms have shown remarkable performance.
These PnP algorithms are similar to traditional iterative algorithms like FISTA, ADMM, or primal-dual splitting (PDS), but differ in that the proximal update is replaced by a call to an application-specific image denoiser, such as BM3D or DnCNN.
The fixed-points of PnP algorithms depend upon an algorithmic stepsize parameter, however, which must be tuned for optimal performance.
In this work, we propose a fast and robust auto-tuning PnP-PDS algorithm that exploits knowledge of the measurement-noise variance that is available from a pre-scan in MRI.
Experimental results show that our algorithm converges very close to genie-tuned performance, and does so significantly faster than existing autotuning approaches. 
\end{abstract}

\IEEEpeerreviewmaketitle

\section{Introduction} \label{sec:intro}

We consider parallel magnetic resonance imaging (MRI) with $C\geq 1$ coils, where the k-space (i.e., Fourier) measurements from the $i$th coil, $\boldsymbol{y}_i\in\mathbb{C}^{M}$, can be expressed as
\begin{align}
	\underbrace{\begin{bmatrix} \boldsymbol{y}_1 \\ \boldsymbol{y}_2 \\ \vdots
			\\ \boldsymbol{y}_C \end{bmatrix}}_{\displaystyle \boldsymbol{y}}
	&=
	\underbrace{\begin{bmatrix}
			\boldsymbol{PFS}_1 \\
			\boldsymbol{PFS}_2 \\
			\vdots\\
			\boldsymbol{PFS}_C\\
	\end{bmatrix}}_{\displaystyle \boldsymbol{A}} \boldsymbol{x}  + \boldsymbol{w} 
\label{eq:y} .
\end{align}
In \eqref{eq:y}, 
$\boldsymbol{x}\in\mathbb{C}^{N}$ is the (vectorized) $N$-pixel image,
$\boldsymbol{S}_i\in\mathbb{C}^{N\times N}$ is a diagonal matrix containing the sensitivity map for the $i^{\text{th}}$ coil, 
$\boldsymbol{F}\in\mathbb{C}^{N\times N}$ is the discrete Fourier transform, 
$\boldsymbol{P}\in\mathbb{R}^{M\times N}$ is a subsampling matrix with $M<N$,
$\boldsymbol{w}\in\mathbb{C}^{CM}$ contains measurement noise of variance $\sigma^2$, 
and 
$\boldsymbol{A}\in\mathbb{C}^{CM\times N}$ is the forward operator.  
Our goal is to recover the image $\boldsymbol{x}$ from measurements $\boldsymbol{y}$ assuming knowledge of $\boldsymbol{A}$.

Often, image recovery is accomplished by posing and then solving an optimization problem of the form
\begin{equation}\label{eq:opti_basic}
	\widehat{\boldsymbol{x}} = \argmin_{\boldsymbol{x}\in \mathbb{C}^{N}} 
	\big\{ \ell(\boldsymbol{A}\boldsymbol{x};\boldsymbol{y}) + \lambda\phi(\boldsymbol{x}) \big\},
\end{equation}
where $\ell(\cdot;\boldsymbol{y})$ is a loss term, $\phi(\cdot)$ is a regularization term, and $\lambda>0$ is a tuning parameter.
Many algorithms have been developed to solve \eqref{eq:opti_basic} in the case that both $\ell(\cdot;\boldsymbol{y})$ and $\phi(\cdot)$ are convex, including 
the alternating direction method of multipliers (ADMM) \cite{Boyd:FTML:11}, 
the fast iterative shrinkage-thresholding algorithm (FISTA) \cite{Beck:JIS:09}, and
primal-dual splitting (PDS) \cite{Chambolle:JMIV:11}.
All three of these algorithms alternate a loss-reducing operation with a so-called ``proximal'' operation of the form
\begin{equation}\label{eq:prox}
	\prox_{\nu\phi}(\boldsymbol{s}) \delequal \argmin_{\boldsymbol{x}} 
	\big\{ \lambda\phi(\boldsymbol{x}) + \tfrac{1}{2\nu} \|\boldsymbol{x} - \boldsymbol{s}\|_2^2 \big\},
\end{equation}
where $\boldsymbol{s}$ and $\nu>0$ are algorithm- and iteration-dependent.

Because \eqref{eq:prox} can be interpreted as maximum a posteriori denoising of $\boldsymbol{s}$ under the prior pdf $p(\boldsymbol{x})\propto e^{-\lambda\phi(\boldsymbol{x})}$, it has been proposed \cite{Venkatakrishnan:GSIP:13,Ono:SPL:17,Sun:TCI:19} to replace \eqref{eq:prox} with a call to a sophisticated application-specific denoiser like BM3D \cite{Dabov:TIP:07} or DnCNN \cite{Zhang:TIP:17}.
In many applications, including MRI \cite{Ahmad:SPM:20}, these ``plug-and-play'' (PnP) image recovery methods methods tend to provide much cleaner recoveries than the traditional optimization approach \eqref{eq:opti_basic}.  

For MRI image recovery, PnP shows similar performance to end-to-end deep neural networks (DNNs) like \cite{Jin:TIP:17,Zbontar:18} but offers several advantages \cite{Ahmad:SPM:20}.
For example, end-to-end DNNs often require a huge corpus of fully sampled k-space data for training, whereas DNN denoisers can be trained using patches from a few images.
Also, the training of end-to-end DNNs is typically dependent on some choice of the forward operator $\boldsymbol{A}$, which can cause performance degradation when the network is applied to recover images measured under a different $\boldsymbol{A}$.
In PnP, the denoiser training does not depend on $\boldsymbol{A}$ and thus there can be no such mismatch error.

In this work, we focus on the PnP-PDS algorithm \cite{Ono:SPL:17}, which alternates the following two steps:
\begin{subequations} \label{eq:PDS}
	\begin{align}
		\boldsymbol{x}_k 
		&= \boldsymbol{f}(\boldsymbol{x}_{k-1} - \gamma_1 \boldsymbol{A}^\mathsf{H}\boldsymbol{v}_{k-1})
		\label{eq:PDS_1}\\
		\boldsymbol{v}_k 
		&=\prox_{\gamma_2 \ell^*}(\boldsymbol{v}_{k-1}+ \gamma_2\boldsymbol{A}(2\boldsymbol{x}_k-\boldsymbol{x}_{k-1}))
		\label{eq:PDS_2} ,
	\end{align}
\end{subequations}
where $\boldsymbol{f}:\mathbb{C}^{N}\rightarrow\mathbb{C}^{N}$ is the denoiser,
$\gamma_1$ and $\gamma_2$ are positive stepsizes, and
$\prox_{\gamma_2 \ell^*}(\boldsymbol{s})=\boldsymbol{s}-\gamma_2\prox_{\gamma_2^{-1}\ell}(\gamma_2^{-1}\boldsymbol{s})$, where $\ell^*(\cdot;\boldsymbol{y})$ denotes the convex conjugate of $\ell(\cdot;\boldsymbol{y})$.
The stepsizes are chosen such that they satisfy $\gamma_1\gamma_2\leq 1/\|\boldsymbol{A}\|_2^2$.
We focus on PnP-PDS since it has advantages over both PnP-FISTA \cite{Sun:TCI:19} and PnP-ADMM \cite{Venkatakrishnan:GSIP:13}:
it yields a first-order algorithm for typical choices of $\ell(\cdot;\boldsymbol{y})$ and does not upper-bound the stepsize $\gamma_1$.

In the common case of quadratic loss $\ell(\boldsymbol{z};\boldsymbol{y})=\frac{1}{2}\|\boldsymbol{z}-\boldsymbol{y}\|_2^2$, the PnP-PDS algorithm \eqref{eq:PDS} becomes
\begin{subequations} \label{eq:PDS_quad}
	\begin{align}
		\boldsymbol{x}_k 
		&= \boldsymbol{f}(\boldsymbol{x}_{k-1} - \gamma_1 \boldsymbol{A}^\mathsf{H}\boldsymbol{v}_{k-1})
		\label{eq:PDS_quad_1}\\
		\boldsymbol{v}_k 
		&\textstyle
                =\frac{1}{1 + \gamma_2}\boldsymbol{v}_{k-1} 
		+ \frac{\gamma_2}{1 + \gamma_2}(\boldsymbol{A}(2\boldsymbol{x}_k - \boldsymbol{x}_{k-1} ) -  \boldsymbol{y})
		\label{eq:PDS_quad_2} .
	\end{align}
\end{subequations}
Rather than solving an optimization problem of the form \eqref{eq:opti_basic}, the PnP-PDS algorithm \eqref{eq:PDS_quad} seeks the solution $\widehat{\boldsymbol{x}}$ to the fixed-point equation
\begin{align}
	\widehat{\boldsymbol{x}} = \boldsymbol{f}\big(\widehat{\boldsymbol{x}}-\gamma_1 \boldsymbol{A}^\mathsf{H}(\boldsymbol{A}\widehat{\boldsymbol{x}}-\boldsymbol{y})\big)
	\label{eq:fixed_point}.
\end{align}
or, equivalently, the solution to a consensus equilibrium problem \cite{Buzzard:JIS:18}.

\begin{figure}[t]
	\centering
	\includegraphics[width = \linewidth]{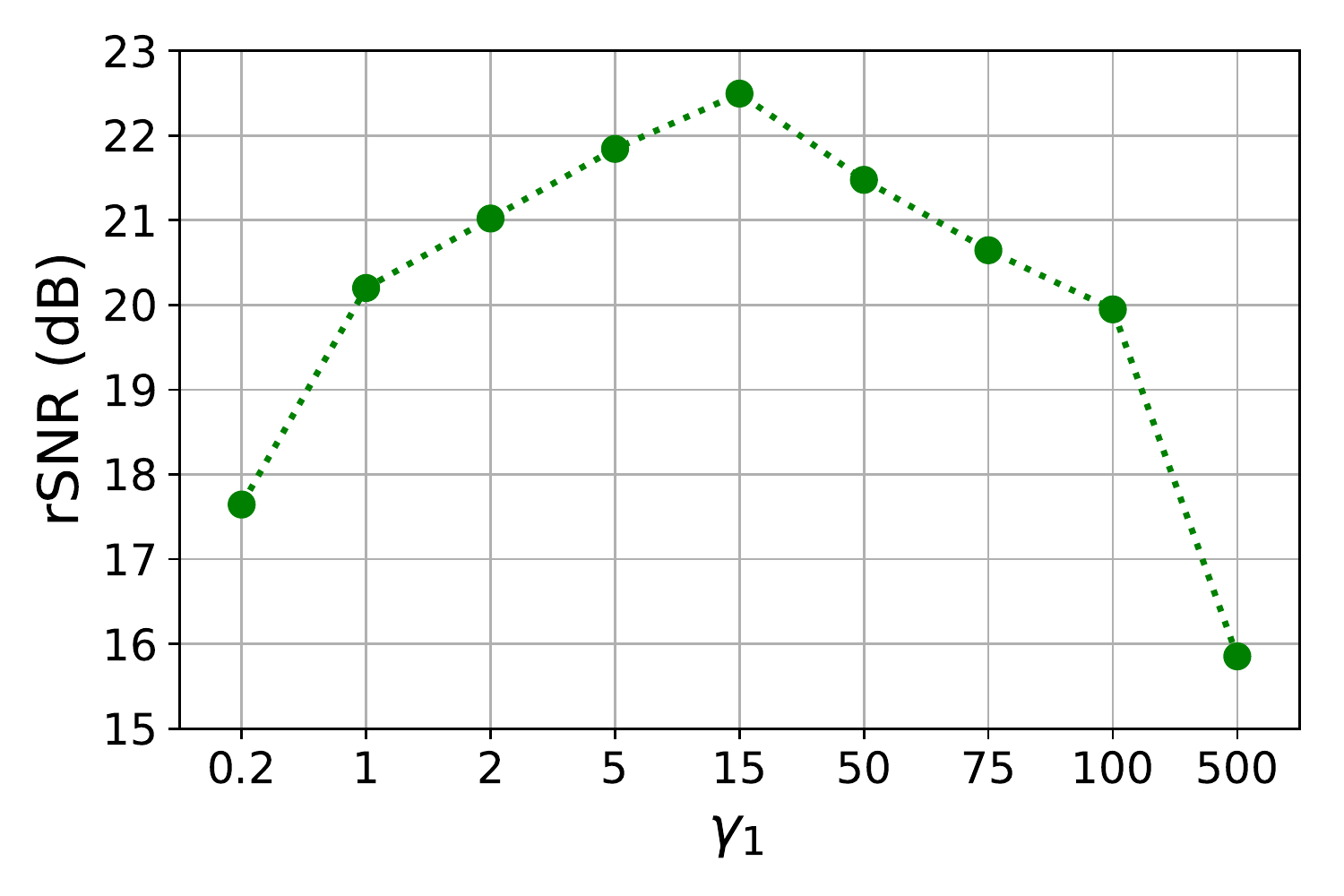}
        \vspace{-5mm}
	\caption{Asymptotic rSNR of quadratic-loss PnP-PDS \eqref{eq:PDS_quad} versus stepsize $\gamma_1$.} 
	\label{fig:PnP_PDS_performance_for_various_gamma_1}
\end{figure}

Equation \eqref{eq:fixed_point} shows that the PnP-PDS fixed point $\widehat{\boldsymbol{x}}$ depends on the choice of the stepsize $\gamma_1$ (but not $\gamma_2$). 
Thus, $\gamma_1$ must be tuned for best recovery performance. 
This is illustrated in Fig.~\ref{fig:PnP_PDS_performance_for_various_gamma_1}, 
which plots asymptotic recovery SNR (rSNR) $\triangleq\frac{\|\boldsymbol{x}\|_2^2}{\|\hat{\boldsymbol{x}} - \boldsymbol{x}\|_2^2}$ versus $\gamma_1$ for a typical image recovery experiment.
Furthermore, the optimal value of $\gamma_1$ is dependent on the forward operator $\boldsymbol{A}$, the noise variance $\sigma^2$, and the image $\boldsymbol{x}$.  
This makes it difficult to tune $\gamma_1$ in practice, when $\boldsymbol{x}$ is unknown.

In this paper, we propose an autotuning variant of PnP-PDS that shows fast convergence and robustness to parameter initializations.
Our method leverages the fact that, in MRI, it is easy to measure the noise variance $\sigma^2$ using a pre-scan.
Before describing our method, we briefly review existing approaches to autotuned PnP.

\section{Existing work}

In the optimization problem \eqref{eq:opti_basic}, it can be difficult to choose the optimal regularization weight $\lambda$.
When $\ell(\cdot;\boldsymbol{y})$ and $\phi(\boldsymbol{x})$ are both convex, \eqref{eq:opti_basic} can be equivalently recast \cite{Lorenz:IP:13} as
\begin{equation}\label{eq:opti_equivalent}
	\widehat{\boldsymbol{x}} = \argmin_{\boldsymbol{x}\in \mathbb{C}^{N}} 
	 \phi(\boldsymbol{x}) \quad \text{subject to}\quad \ell(\boldsymbol{A}\boldsymbol{x};\boldsymbol{y}) \leq \mu,
\end{equation}
for some $\lambda$-dependent parameter $\mu>0$.
The formulation \eqref{eq:opti_equivalent} may be preferable to \eqref{eq:opti_basic} in that $\mu$ may be easier to choose than $\lambda$.
For example, with quadratic loss $\ell(\boldsymbol{z};\boldsymbol{y})=\frac{1}{2}\|\boldsymbol{z}-\boldsymbol{y}\|_2^2$ and $\sigma^2$-variance additive noise $\boldsymbol{w}$, the value of $\ell(\boldsymbol{z};\boldsymbol{y})$ at the optimal setting $\boldsymbol{z}=\boldsymbol{Ax}$ concentrates at $CM\sigma^2/2$ (as $CM$ grows large, by the law of large numbers), motivating the choice of 
\begin{align}
\mu \approx CM\sigma^2/2
\label{eq:mu}
\end{align}
in \eqref{eq:opti_equivalent} when the noise variance $\sigma^2$ is known.
This idea has been routinely applied for image recovery (see, e.g., \cite{Hunt:TCOMP:73,Galatsanos:TIP:92}) and is sometimes referred to as the discrepancy principle \cite{Morozov:Book:12}.

Although the above formulation pertains to the optimization approach to image recovery, similar ideas have appeared in the PnP literature, as we now describe.

\subsection{PDS-ATO}

Inspired by Morozov's discrepancy principle \eqref{eq:opti_equivalent}-\eqref{eq:mu}, Ono \cite{Ono:SPL:17} proposed PnP-PDS \eqref{eq:PDS} with an indicator loss of the form
\begin{align}
	\ell(\boldsymbol{z};\boldsymbol{y}) 
	&= \begin{cases}
		0 & \frac{1}{CM}\|\boldsymbol{y}-\boldsymbol{z}\|^2 \leq \beta\sigma^2 \\
		\infty & \text{otherwise}, \\
	\end{cases}
\end{align}
where $\beta\approx 1$ is a user-adjustable parameter 
(we fix $\beta=0.95$ in all of our experiments).
This yields the iteration
\begin{subequations}\label{eq:PDS_ono}
	\begin{align}
		\boldsymbol{x}_{k} 
		&= \boldsymbol{f}(\boldsymbol{x}_{k-1} - \gamma_1\boldsymbol{A}^{\mathsf{H}}\boldsymbol{v}_{k-1})
		\label{eq:PDS_ono_1}\\
		\boldsymbol{v}_{k} 
		&= \text{max}\left\{0, 1 - \frac{\gamma_2\sqrt{\beta CM}\sigma}{\|\boldsymbol{v}_{k-1} + \gamma_2\boldsymbol{A}(2\boldsymbol{x}_{k} - \boldsymbol{x}_{k-1}) - \gamma_2\boldsymbol{y}\|_2} \right\}\nonumber \\ & \qquad \qquad  \quad (\boldsymbol{v}_{k-1} + \gamma_2\boldsymbol{A}(2\boldsymbol{x}_{k} - \boldsymbol{x}_{k-1}) - \gamma_2\boldsymbol{y})
		\label{eq:PDS_ono_2} ,
	\end{align}
\end{subequations}
where it is typical to choose $\gamma_2=1/(\gamma_1\|\boldsymbol{A}\|_2^2)$ for stability.
We will refer to \eqref{eq:PDS_ono} as the AutoTune-Ono (ATO) approach to PDS, i.e., PDS-ATO.
It should be emphasized that, in \eqref{eq:PDS_ono}, the stepsize parameters $\gamma_1,\gamma_2$ affect convergence speed but \emph{not} the fixed points.
As we will see, \eqref{eq:PDS_ono} has good fixed points but somewhat slow convergence speed, regardless of the choice of $\gamma_1$.

\subsection{PDS-ATM}

In the fixed-point expression \eqref{eq:fixed_point} for quadratic-loss PnP-PDS, it can be seen that $\gamma_1$ magnifies the residual measurement-error $\boldsymbol{y}-\boldsymbol{A}\widehat{\boldsymbol{x}}$.
Thus, a larger value of $\gamma_1$ should correspond to a smaller value of $\|\boldsymbol{y}-\boldsymbol{A}\widehat{\boldsymbol{x}}\|_2^2$, and vice versa.
This observation suggests that Morozov's discrepancy principle can somehow be used to adapt $\gamma_1$ at each iteration in quadratic-loss PnP-PDS. 
For example, if $\|\boldsymbol{y}-\boldsymbol{A}\widehat{\boldsymbol{x}}\|_2^2$ is larger than the target value at a given iteration, then $\gamma_1$ could be increased.
A first incarnation of this approach was proposed by Liu et al.\ in \cite{Liu:ISMRM:20}:
\begin{subequations}\label{eq:PDS-ATM}
	\begin{align}
		\boldsymbol{x}_{k} 
		&= \boldsymbol{f}(\boldsymbol{x}_{k-1} - \gamma_{1,k-1} \boldsymbol{A}^\mathsf{H}\boldsymbol{v}_{k-1}) 
		        \label{eq:ATM_quad_1}\\
		\boldsymbol{v}_{k} 
		&\textstyle
                 = \frac{1}{1 + \gamma_{2,k-1}}\boldsymbol{v}_{k-1} + \frac{\gamma_{2,k-1}}{1 + \gamma_{2,k-1}}(\boldsymbol{A}(2\boldsymbol{x}_k - \boldsymbol{x}_{k-1} ) -  \boldsymbol{y})
		        \label{eq:ATM_quad_2}\\
		\gamma_{1,k} &= \gamma_{1,k-1}\frac{\|\boldsymbol{y}- \boldsymbol{A}\boldsymbol{x}_k\|_2^2}{\beta CM\sigma^2} 
                        \label{eq:ATM_mul_update}\\
		\gamma_{2,k} &= \frac{1}{\gamma_{1,k}\|\boldsymbol{A}\|_2^2}
                        \label{eq:ATM_gam2_update}.
	\end{align}
\end{subequations}
We will refer to \eqref{eq:PDS-ATM} as PDS-ATM. 
Note that the multiplicative update step \eqref{eq:ATM_mul_update} causes $\gamma_1$ to be increased when $\frac{1}{CM}\|\boldsymbol{y}-\boldsymbol{Ax}_k\|^2 >\beta\sigma^2$ and decreased when $\frac{1}{CM}\|\boldsymbol{y}-\boldsymbol{Ax}_k\|^2 <\beta\sigma^2$, with the goal of driving $\gamma_1$ to a value for which $\frac{1}{CM}\|\boldsymbol{y}-\boldsymbol{A}\widehat{\boldsymbol{x}}\|^2 =\beta\sigma^2$.

\subsection{PDS-ATM1}

\begin{figure}[t]
	\centering
	\includegraphics[width = \linewidth]{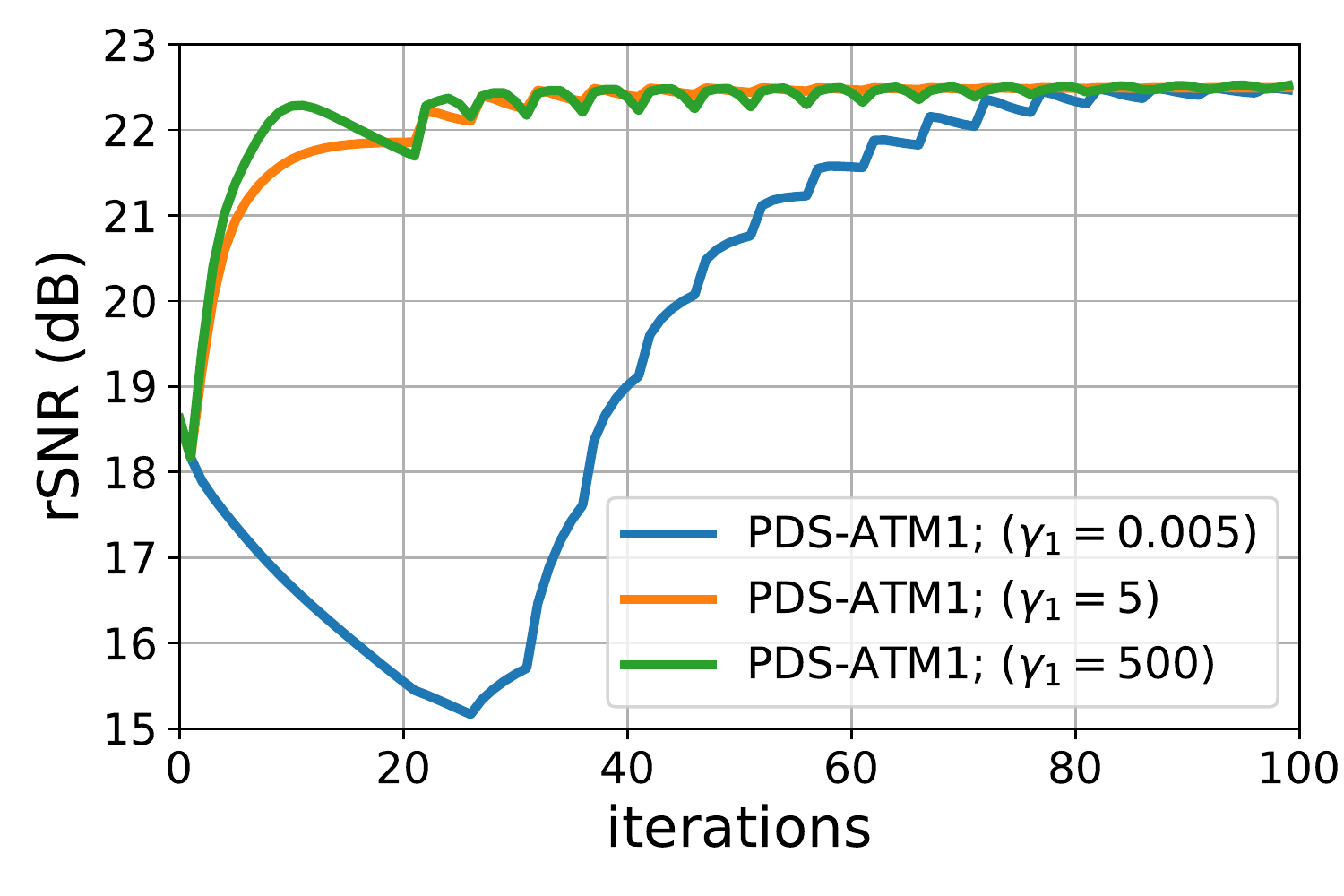}
        \vspace{-7mm}
	\caption{rSNR versus iteration for PDS-ATM1 under several initial $\gamma_{1}$.}
	\label{fig:PDS-ATM1_for_various_gamma_1_0}
\end{figure}

To help promote stability, the authors of \cite{Liu:ISMRM:20} made some modifications to \eqref{eq:PDS-ATM} for practical implementation. 
They are summarized in Alg.~\ref{alg:PDS-ATM1}, which we shall refer to as PDS-ATM1.
In Alg.~\ref{alg:PDS-ATM1}, the stepsizes $\gamma_{1,k}$ and $\gamma_{2,k}$ are updated only once every $5$ iterations, starting from iteration $k=20$. 
Although these modifications enhance the stability of PDS-ATM1, they slow down its convergence, 
as seen in Fig.~\ref{fig:PDS-ATM1_for_various_gamma_1_0}.  
And even with these modifications, we find that Alg.~\ref{alg:PDS-ATM1} fails to converge when the initial $\gamma_{1}$ is too large.
These shortcoming motivate our approach to autotuned PnP-PDS, which is detailed in the next section.

\begin{algorithm}[t]
	\caption{The PDS-ATM1 algorithm \cite{Liu:ISMRM:20}}
	\begin{algorithmic}[1]
		\Require  $\gamma_1>0, \boldsymbol{x}_0 \in \mathbb{C}^{N}, \boldsymbol{v}_0 \in \mathbb{C}^{N}, \beta\in(0,1]$
		\State $\gamma_{1,0}=\gamma_1$
		\State $\gamma_{2,0} = \frac{1}{\gamma_{1,0}\|\boldsymbol{A}\|_2^2}$
		\For {$k = 1,2,3, \ldots $}
		\State $ \boldsymbol{x}_k = \boldsymbol{f}(\boldsymbol{x}_{k-1} - \gamma_{1,k-1} \boldsymbol{A}^\mathsf{H}\boldsymbol{v}_{k-1}) $
		\State $ \boldsymbol{v}_k = \frac{1}{1 + \gamma_{2,k-1}}\boldsymbol{v}_{k-1} + \frac{\gamma_{2,k-1}}{1 + \gamma_{2,k-1}}(\boldsymbol{A}(2\boldsymbol{x}_k - \boldsymbol{x}_{k-1} ) -  \boldsymbol{y}) $
		\If{$k>19$ \textbf{and} mod$(k,5) == 0$}\label{line:ATM1_modification}
		\State $\gamma_{1,k} = \gamma_{1,k-1}\frac{\|\boldsymbol{y}- \boldsymbol{A}\boldsymbol{x}_k\|_2^2}{\beta CM\sigma^2} $\label{line:ATM1_mul_update}
		\State $\gamma_{2,k} = \frac{1}{\gamma_{1,k}\|\boldsymbol{A}\|_2^2}$
		\Else
		\State $\gamma_{1,k} = \gamma_{1,k-1}$
		\State $\gamma_{2,k} = \gamma_{2,k-1}$
		\EndIf
		\EndFor 
	\end{algorithmic}\label{alg:PDS-ATM1}
\end{algorithm}

\section{Proposed Algorithm: PDS-ATM2} \label{sec:proposed_algo}

The proposed method, summarized in Alg.~\ref{alg:PDS-ATM2}, is called PDS-ATM2.
It builds on the multiplicative-update idea from PDS-ATM \eqref{eq:PDS-ATM} but adds two key refinements: damping and restarting.  
These two refinements help PDS-ATM2 to converge quickly and maintain robustness to the choice of initial $\gamma_{1}$.
A detailed explanation of PDS-ATM2 follows.

\begin{algorithm}[t]
	\caption{The PDS-ATM2 algorithm (proposed)}
	\begin{algorithmic}[1]
		\Require $\gamma_1>0, \boldsymbol{x}_0 \in \mathbb{C}^{N}, \boldsymbol{v}_0 \in \mathbb{C}^{N}, \beta\in(0,1],  \alpha\in(0,1]$ 
                \State \restartnow $=$ FALSE, \restartallowed $=$ TRUE
		\State $\gamma_{1,0}=\gamma_1$
		\State $\gamma_{2,0} = \frac{1}{\gamma_{1,0}\|\boldsymbol{A}\|_2^2}$
		\For {$k = 1,2,3, \ldots $}
		\State $ \boldsymbol{x}_k = \boldsymbol{f}(\boldsymbol{x}_{k-1} - \gamma_{1,k-1} \boldsymbol{A}^\mathsf{H}\boldsymbol{v}_{k-1}) $ \label{line:x_k_update}
		\State $ \boldsymbol{v}_k = \frac{1}{1 + \gamma_{2,k-1}}\boldsymbol{v}_{k-1} + \frac{\gamma_{2,k-1}}{1 + \gamma_{2,k-1}}(\boldsymbol{A}(2\boldsymbol{x}_k - \boldsymbol{x}_{k-1} ) -  \boldsymbol{y})$\label{line:v_k_update} 

		\If{$\|\boldsymbol{y}- \boldsymbol{A}\boldsymbol{x}_k\|_2^2 > \|\boldsymbol{y}- \boldsymbol{A}\boldsymbol{x}_{k-1}\|_2^2$}\label{line:rst_now_start}
		\State \restartnow $=$ TRUE
		\Else
		\State \restartnow $=$ FALSE
		\EndIf
                \label{line:rst_now_end}

		\If{$\|\boldsymbol{y}- \boldsymbol{A}\boldsymbol{x}_k\|_2^2 < {\beta CM\sigma^2}$}\label{line:rst_allow_begin} 
		\State \restartallowed $=$ FALSE
		\ElsIf{$\|\boldsymbol{y}- \boldsymbol{A}\boldsymbol{x}_k\|_2^2 > {1.1\beta CM\sigma^2}$}
                \label{line:rst_upper_tresh} 
		\State \restartallowed $=$ TRUE
		\EndIf\label{line:rst_allow_end} 

		\If{\restartallowed \textbf{and} \restartnow}
                \label{line:rst_control}
		\State $\gamma_{1,k} = \gamma_{1,0}$
                \label{line:rst_gamma_1}
		\Else
		\State $\gamma_{1,k} = \alpha\gamma_{1,k-1}\frac{\|\boldsymbol{y}- \boldsymbol{A}\boldsymbol{x}_k\|_2^2}{\beta CM\sigma^2} + (1-\alpha)\gamma_{1,k-1}$ 
                \label{line:ATM2_mul_update}
		\EndIf

		\State $\gamma_{2,k} = \frac{1}{\gamma_{1,k}\|\boldsymbol{A}\|_2^2}$\label{line:rst_gamma_2}
                \label{line:ATM2_gam2_update}

		\If{$k>2$ \textbf{and} $\gamma_{1,k}==\gamma_{1,k-1}==\gamma_{1,k-2}$}\label{line:multiple_rst_start}
		\State $\gamma_{1,0} = 10\gamma_{1,0}$
		\EndIf\label{line:multiple_rst_end}
		\EndFor 
	\end{algorithmic}
	\label{alg:PDS-ATM2}
\end{algorithm}

Because PDS-ATM2 builds on PDS-ATM, lines~\ref{line:x_k_update}--\ref{line:v_k_update} of Alg.~\ref{alg:PDS-ATM2} are identical to \eqref{eq:ATM_quad_1} and \eqref{eq:ATM_quad_2}, respectively. 
These two lines mirror the main update steps of quadratic-loss PnP-PDS, \eqref{eq:PDS_quad_1} and \eqref{eq:PDS_quad_2}. 

Skipping ahead, line~\ref{line:ATM2_mul_update} of Alg.~\ref{alg:PDS-ATM2} performs the (damped) multiplicative update of $\gamma_1$.
If the damping parameter $\alpha$ was $=1$, there would be no damping, and the PDS-ATM2 update of $\gamma_1$ in line~\ref{line:ATM2_mul_update} would be identical to the PDS-ATM update of $\gamma_1$ in \eqref{eq:ATM_mul_update}.
But when $\alpha\in(0,1)$, line~\ref{line:ATM2_mul_update} slows the update of $\gamma_1$.
Importantly, the form of line~\ref{line:ATM2_mul_update} ensures that the fixed-point value of $\gamma_1$ is unaffected by the choice of $\alpha\in(0,1]$.
PDS-ATM2 then updates the value of $\gamma_2$ in line~\ref{line:ATM2_gam2_update} of Alg.~\ref{alg:PDS-ATM2} in the same way that PDS-ATM updates it in \eqref{eq:ATM_gam2_update}.


As visible from lines~\ref{line:rst_now_start}--\ref{line:rst_now_end} of Alg.~\ref{alg:PDS-ATM2}, the \restartnow flag is raised when the residual measurement error $\|\boldsymbol{y}-\boldsymbol{Ax}_k\|_2^2$ increases from one iteration to the next.
This is because residual error is expected to start large and decrease over the iterations; 
an increasing residual error is a sign that the value of $\gamma_1$ has become misadjusted and should be reset, as in line~\ref{line:rst_gamma_1}.

But there is an exception: we want the residual measurement error $\|\boldsymbol{y}-\boldsymbol{Ax}_k\|_2^2$ to increase when it lies below the target value of $\beta CM\sigma^2$.
When this situation is detected, in line~\ref{line:rst_allow_begin}, the \restartallowed flag is set to FALSE. 
If the residual measurement error grows sufficiently above the target value, as in line~\ref{line:rst_upper_tresh}, the \restartallowed flag is restored to TRUE.


Finally, a very small $\gamma_{1}$ (re)initialization could cause the algorithm to continually restart.  
This issue is circumvented by lines~\ref{line:multiple_rst_start}--\ref{line:multiple_rst_end} of Alg.~\ref{alg:PDS-ATM2}, which increase the value of $\gamma_{1,0}$.

\section{Numerical Experiments}\label{sec:results}

We evaluated our method using $10$ mid-slice, non-fat-suppressed, $128\times128$, test images from the single-coil fastMRI knee image database~\cite{Zbontar:18}. 
To construct the multi-coil measurements $\boldsymbol{y}$, we simulated 4 receive coils using the Biot-Savart law and performed k-space subsampling using a fixed Cartesian mask with acceleration $4$ (i.e. $\frac{N}{M} = 4$).  
White Gaussian noise was then added to obtain images with average signal-to-noise ratio (SNR) in $\{15,17,20,23\}$ dB.   

The proposed PDS-ATM2 algorithm was compared to PDS-ATM1, PDS-ATO, the U-Net from \cite{Zbontar:18}, and genie-tuned PnP-PDS. 
By ``genie-tuned PnP-PDS,'' we mean PnP-PDS with the rSNR-maximizing value of $\gamma_1$ (found by grid-search).
Genie-tuned PnP-PDS is not a practical algorithm, but an upper bound on what one could expect from optimally autotuned version of PnP-PDS. 

All image reconstruction methods used an $\boldsymbol{A}$ constructed with sensitivity maps $\boldsymbol{S}_i$ estimated from $\boldsymbol{y}$ using ESPIRiT~\cite{Uecker:MRM:14}.  
For the denoiser, a modified DnCNN was trained on MRI knee images that were different from the test images; see \cite{Ahmad:SPM:20} for details.  
The U-Net was trained separately at each noise level using the same set of training images used for the denoiser.
All autotuned-PnP methods used $\beta=0.95$, and PDS-ATM2 used the damping value $\alpha = 0.2$. 

\begin{figure*}[t]
	\centering
	\begin{tabular}{ccc}
		\subfigure[]{\includegraphics[clip,width = 0.32\linewidth]{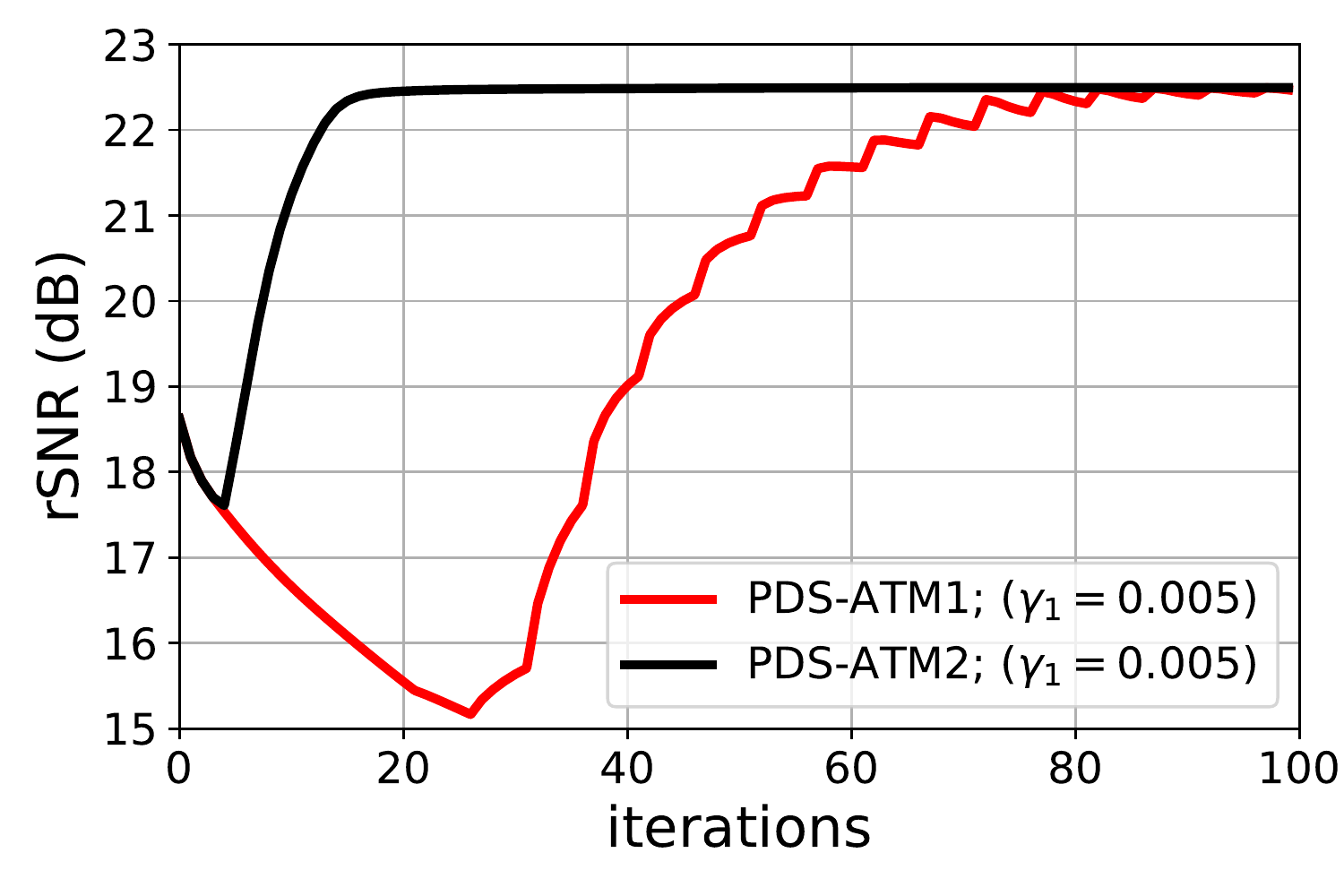}}
		\subfigure[]{\includegraphics[clip,width = 0.32\linewidth]{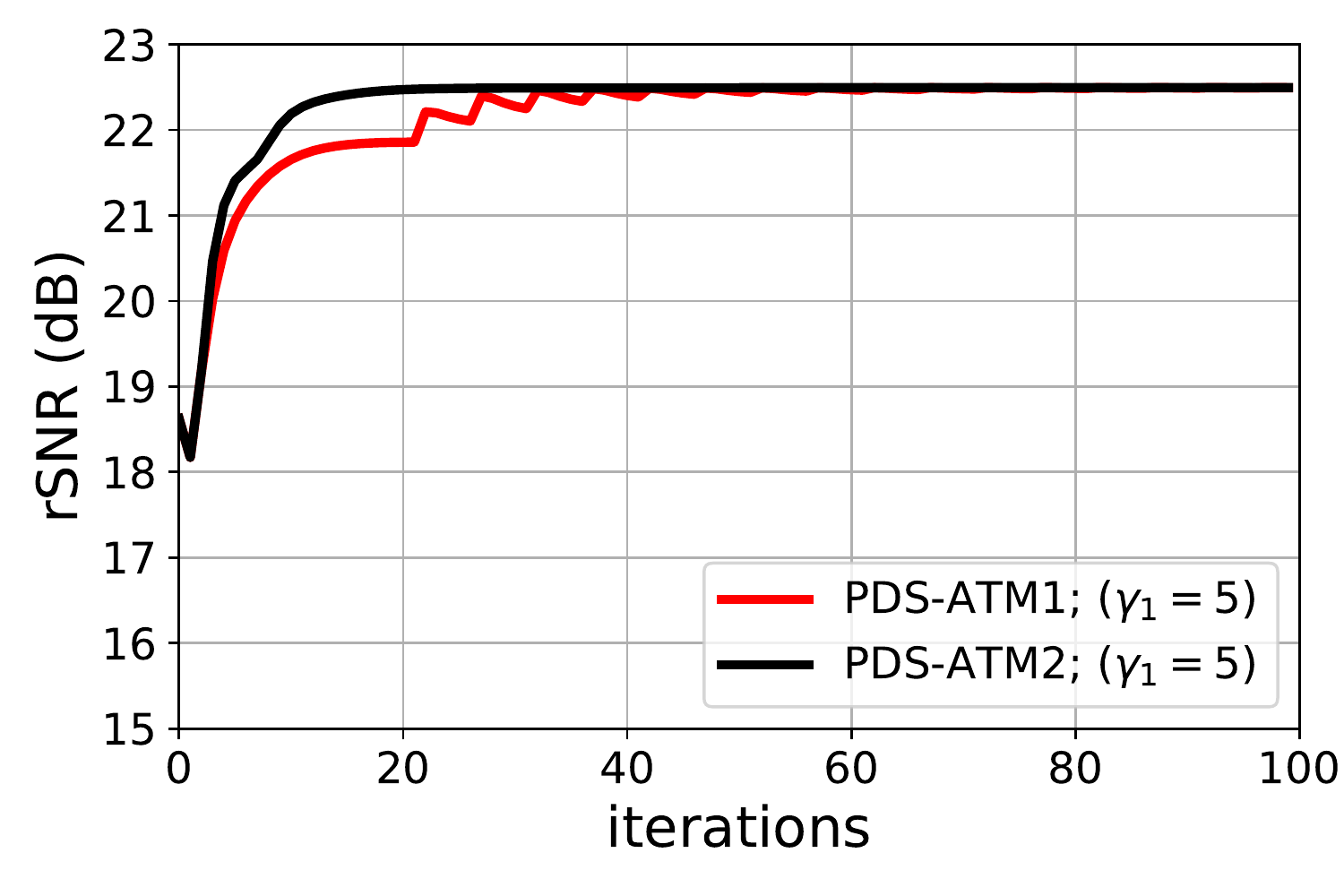}}
		\subfigure[]{\includegraphics[width = 0.32\linewidth]{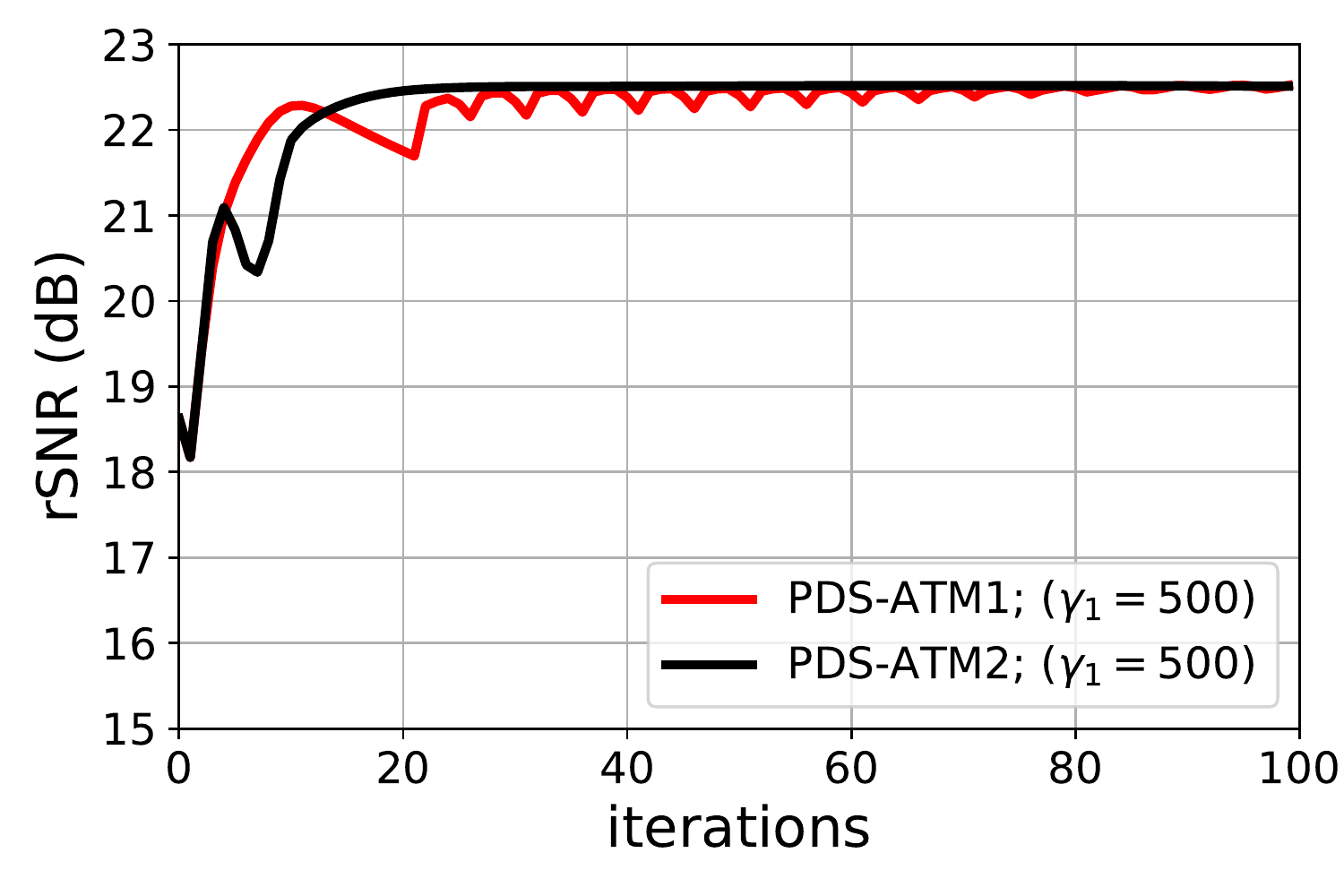}}
	\end{tabular}
        \vspace{-3mm}
	\caption{rSNR versus iteration for PDS-ATM1 and PDS-ATM2 under various choices of initial $\gamma_{1}$.}
	\label{fig:ATM_1_2_comparison}
\end{figure*}

Figure~\ref{fig:ATM_1_2_comparison} shows rSNR versus iteration for PDS-ATM1 and PDS-ATM2 for several choices of initial $\gamma_1$.
From the figure, we see that PDS-ATM2 converges much faster and smoother than PDS-ATM1 for all choices of initial $\gamma_{1}$.
For this reason, we consider only PDS-ATM2 in the sequel.

\begin{figure}[t]
	\centering
	\includegraphics[width = \linewidth]{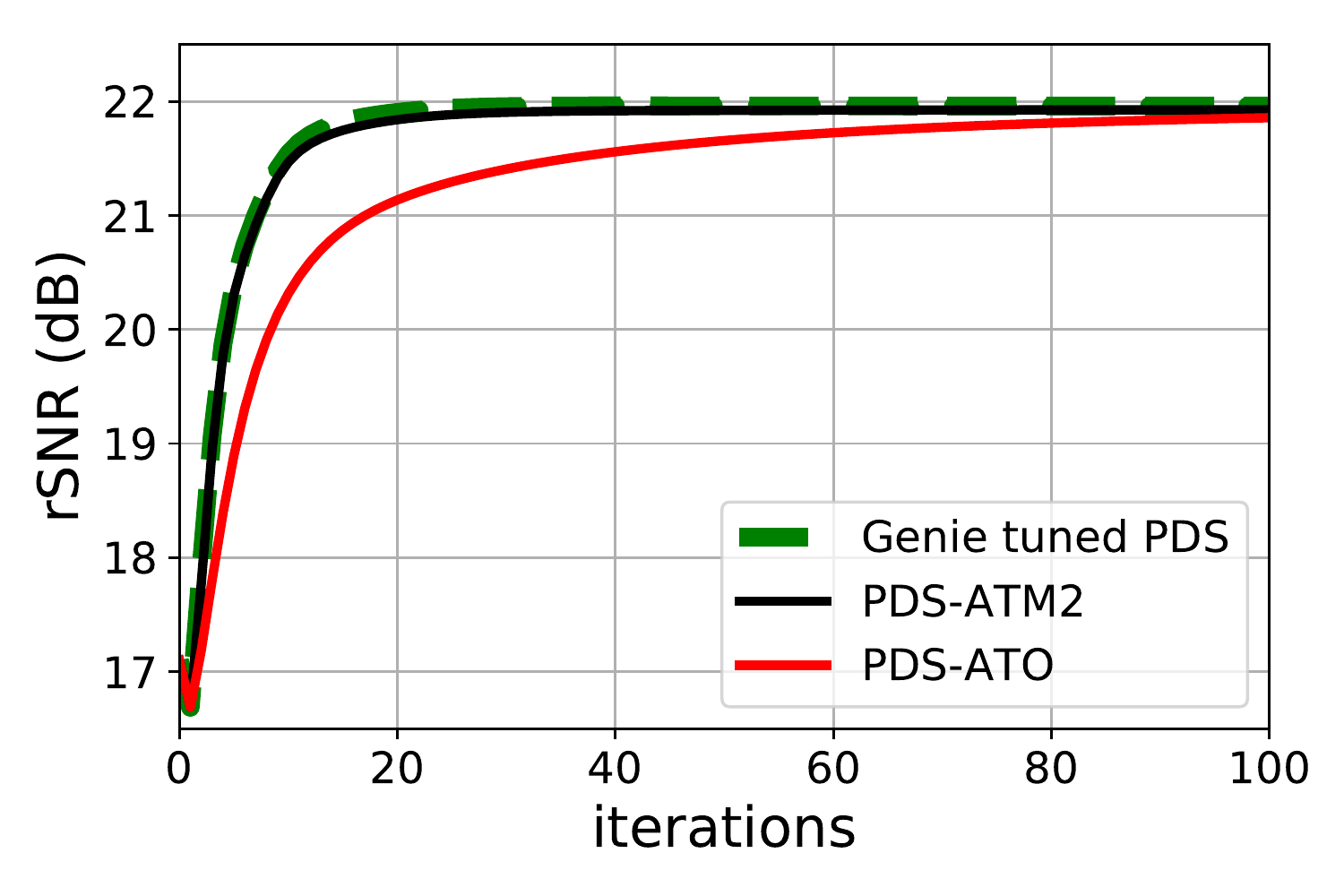}
        \vspace{-7mm}
	\caption{rSNR versus iteration for several PnP algorithms}
	\label{fig:convergence_speed_comparison}
\end{figure}

Figure~\ref{fig:convergence_speed_comparison} shows rSNR versus iteration for PDS-ATM2, PDS-ATO, and genie-tuned PDS.
There we see that PDS-ATM2 converges as fast as genie-tuned PDS (i.e., $<30$ iterations) and significantly faster than PDS-ATO.  

\begin{figure}[t]
	\includegraphics[width = \linewidth]{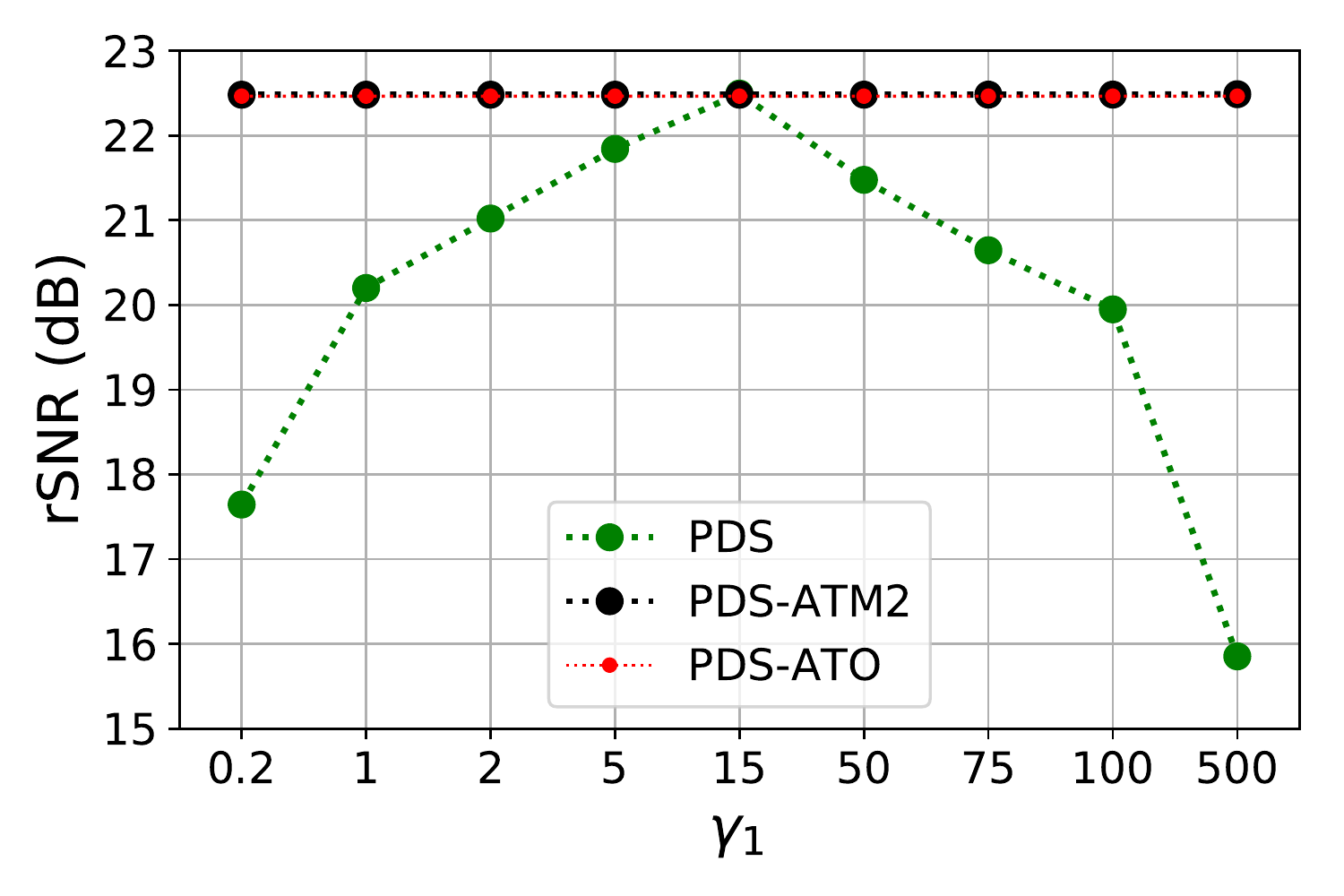}
        \vspace{-7mm}
	\caption{Asymptotic rSNR versus (initial) $\gamma_1$ for several PnP algorithms.}
	\label{fig:Autotune_comparison_ATO_ATM2}
\end{figure}

Figure~\ref{fig:Autotune_comparison_ATO_ATM2} shows rSNR versus (initial) $\gamma_1$ for PDS-ATM2, PDS-ATO, and PDS.
It shows that standard PnP-PDS is very sensitive to the value of $\gamma_1$, while PDS-ATO and PDS-ATM2 are not, as a consequence of autotuning.

Table~\ref{tab:comparison} shows the asymptotic values of rSNR and SSIM, averaged over the $10$ test images, for the methods under consideration.
From the table, we see that the average asymptotic-performances of PDS-ATO and PDS-ATM2 are identical and very close to that of genie-aided PnP-PDS, which plays the role of an impractical upper bound.
The table also shows that the genie-tuned and autotuned PnP approaches give slightly better average rSNR than the U-Net, but slightly worse average SSIM;
overall, their average performance is similar.

\begin{figure*}[t]
	\centering
	\includegraphics[width = \linewidth]{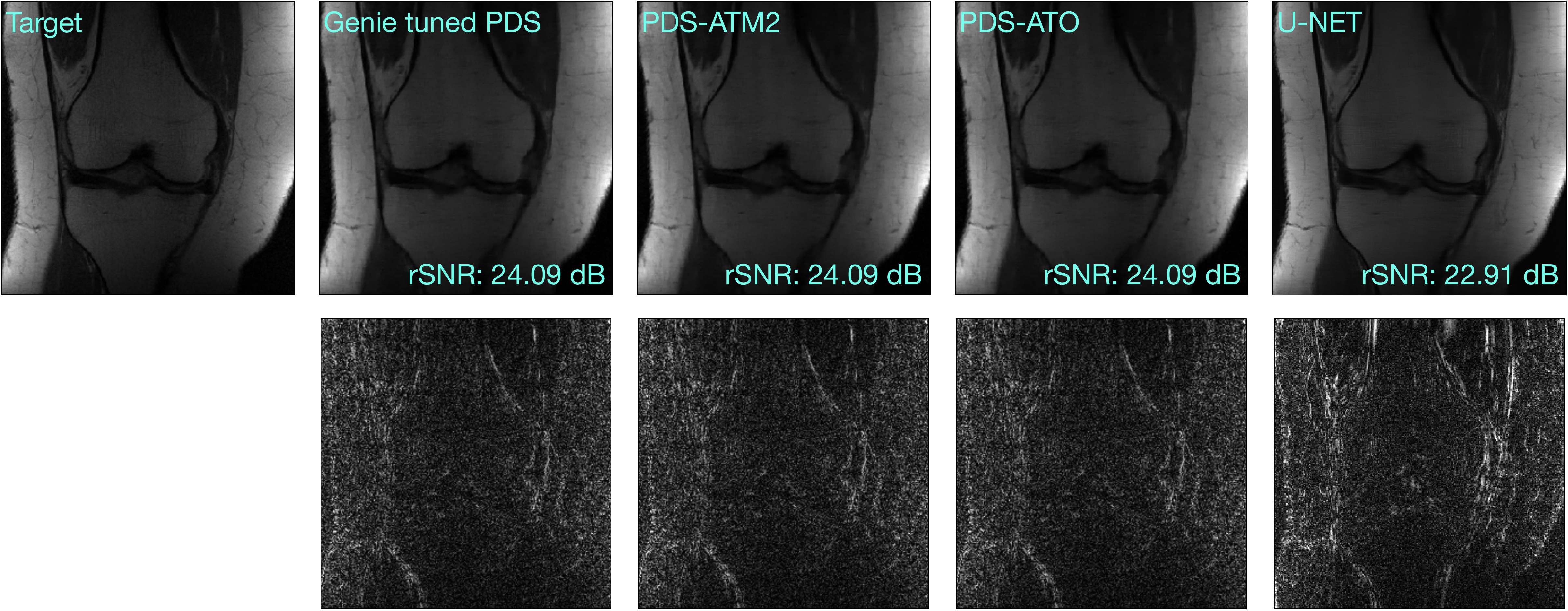}
        \vspace{-5mm}
	\caption{Example recovery of a knee image from the fastMRI dataset.  The top row shows the ground-truth image and several images recoveries with their corresponding rSNR. The bottom row shows the corresponding error images $\times2.4$.}
	\label{fig:recon_comparison}
\end{figure*}

Figure~\ref{fig:recon_comparison} shows an example test image, several recoveries, and their corresponding error images.
For this example, the three PnP methods performed nearly identically and noticeably better than the U-Net (whose rSNR was $1.18$ dB worse).

\begin{table*}[t]
	\centering
	\caption{rSNR and SSIM averaged over the 10 test images for several measured SNRs}
	\begin{tabular}{ | c|| c| c|| c| c|| c| c|| c| c|}
		\hline
		\multicolumn{1}{|c||}{} &\multicolumn{2}{c||}{Avg meas. SNR: 15 dB}  &\multicolumn{2}{c||}{Avg meas. SNR: 17 dB} &\multicolumn{2}{c||}{Avg meas. SNR: 20 dB} &\multicolumn{2}{c|}{Avg meas. SNR: 23 dB}\\
		\cline{2-9}
		\multicolumn{1}{|c||}{} & \multicolumn{1}{c|}{rSNR(dB)} & \multicolumn{1}{c||}{SSIM} & \multicolumn{1}{c|}{rSNR(dB)} & \multicolumn{1}{c||}{SSIM} & \multicolumn{1}{c|}{rSNR(dB)} & \multicolumn{1}{c||}{SSIM} & \multicolumn{1}{c|}{rSNR(dB)} & \multicolumn{1}{c|}{SSIM}\\
		\hline
		\hline
		Genie tuned PDS & 21.02 &   0.804&  21.42&   0.820 &  21.96&   0.839 &  22.42&   0.854   \\\hline
		\hline
		PDS-ATM2 & \textbf{21.00} & 0.805  & \textbf{21.44}&  0.817 & \textbf{21.93} &  0.833 & \textbf{22.33}&  0.850  \\
		\hline
		PDS-ATO& \textbf{21.00} &0.805& \textbf{21.44}&   0.817 & \textbf{21.93}&   0.833 & \textbf{22.33}&   0.850  \\
		\hline
		U-Net&20.77 & \textbf{0.838} & 21.07& \textbf{0.845} & 21.34& \textbf{0.841} & 21.50& \textbf{0.859} \\
		\hline
	\end{tabular}
	\label{tab:comparison}
\end{table*}

\section{Conclusions}

In this paper, we considered parallel MRI image recovery using the PnP framework.  
We focused on the PnP-PDS algorithm, since it is both flexible and computationally efficient.  
We first showed that PnP performance is strongly dependent on the choice of the stepsize, $\gamma_1$, motivating the careful tuning of this parameter. 
We then reviewed several existing methods that circumvent this tuning issue using Morozov's discrepancy principle, which says that the residual measurement error in each k-space sample should be close to the measurement noise variance, $\sigma^2$, which can be measured using an MRI pre-scan.
We showed that, although the existing methods eventually converge to a good recovery, they converge somewhat slowly.
Therefore, we proposed a new autotuning PnP-PDS algorithm that converges quickly and maintains robustness to the choice of initial $\gamma_1$.  
Through numerical experiments with fastMRI knee images, we showed that the performance of our proposed technique is very close to genie-tuned PnP-PDS, and very close to the U-Net, a recently proposed end-to-end deep-learning method.

\section*{Acknowledgment}
The authors thank Edward T. Reehorst for sharing the trained Dn-CNN denoiser and the U-Net training code.

\bibliographystyle{IEEEtran}
\bibliography{macros_abbrev,misc,sparse,machine,books}
\end{document}